\documentstyle[aps,pre,epsfig]{revtex}

\begin{document}
\draft
\title{{\bf Electromagnetic modes in cold magnetized strongly coupled plasmas}}
\author{I.M. Tkachenko $^{a)}$, J. Ortner $^{b)}$, and V.M. Rylyuk $^{c)}$}
\address{$^{a)}${\it Departament de Matematica Aplicada, ETSII, 
Universitat Politecnica, Cami de Vera s/n, Apt. 22012 E-46071 Spain}\\
$^{b)}${\it Institut f\"ur Physik, Humboldt Universit\"{a}t zu Berlin, 
Invalidenstr. 110, D-10115 Berlin, Germany}\\
$^{c)}${\it Department of Theoretical Physics, University of Odesa, 
Dvorjanskaja 2, 270100, Odesa, Ukraine}}

\date{}
\maketitle

\begin{abstract}
The spectrum of electromagnetic waves propagating in a strongly coupled
magnetized fully ionized hydrogen plasma is found.{\ The ion motion and
damping being neglected, the influence of the Coulomb coupling on the
electromagnetic spectrum is analyzed.}
\end{abstract}


\section{Introduction}

The aim of this paper is to find the spectrum of electromagnetic waves
propagating in a strongly coupled magnetized fully ionized hydrogen plasma 
\cite{conf} without taking into account the ion motion. We make use of the
dielectric tensor of cold magnetized plasmas constructed in Ref.{\cite
{Adamyan} by means} of the classical theory of moments{.}

In neglect of thermal motion this dielectric tensor reads: 
\begin{equation}
\varepsilon _{\mu \nu }=\left( 
\begin{array}{ccc}
\varepsilon _{\perp } & ig & 0 \\ 
-ig & \varepsilon _{\perp } & 0 \\ 
0 & 0 & \varepsilon _{\Vert }
\end{array}
\right) ,  \label{3.0}
\end{equation}
where $\varepsilon _{\perp }$ and $\varepsilon _{\Vert }$ are the transverse
and longitudinal (with respect to the external magnetic field, $\vec{B}$)
components of the tensor.

We will consider the damping of the modes in question as negligibly small.
This assumption obviously can be verified only experimentally. The damping 
can be essential and must be taken into account near the cyclotron
resonances. Here the thermal motion of the particles leading to the spatial
dispersion must be accounted for also. Thus our results are valid only far
from the cyclotron resonances and in coupled plasma systems with the plasma 
parameter $\Gamma =e^{2}/aT\gtrsim 1$ ( -e is the electron charge, $a$ is the 
Wigner-Seitz radius and $T$ is the 
plasma temperature). For laboratory plasmas this condition implies the 
temperature $T\sim 2-3 \, eV$ and the number density of electrons 
$n\gtrsim 10^{21}cm^{-3}$. The electrical conductivity $\sigma $ of such 
systems with strong Coulomb coupling is of the order of $\omega _{p}$, so 
that their effective collisional frequency $\nu =\omega _{p}^{2}/4\pi \sigma $
 is at least an order of magnitude smaller than $\omega _{p}$.
Similar conditions can  also be realized in 
astrophysical systems (crust of neutron stars, the interior of white dwarfs 
and large planets with very strong magnetic fields, etc.).
Further we will regard only long wavelengths modes for which the condition 
of a cold plasma \cite{A} holds. In addition the frequencies
of these modes will be presumed to be much higher than the ion cyclotron
frequency, so that the ion motion contribution could be neglected.

The components of the dielectric tensor of a system under consideration were
found in{\cite{Adamyan}}, and within the first approximation in the ratio $%
\sqrt{m/M}$ ($m$ and $M$ being the electron and the ion masses): 
\begin{equation}
\varepsilon _{\perp }=1-\omega _{p}^{2}\,\frac{\omega ^{2}-\Omega _{\perp
}^{2}}{(\omega ^{2}-\Omega _{\perp }^{2})^{2}-\omega ^{2}\omega _{B}^{2}}%
,\quad \varepsilon _{\Vert }=1-\frac{\omega _{p}^{2}}{\omega ^{2}-\Omega
_{\Vert }},\quad g=\omega _{p}^{2}\frac{\omega \omega _{B}}{(\omega
^{2}-\Omega _{\perp }^{2})^{2}-\omega ^{2}\omega _{B}^{2}},  \label{ecomp}
\end{equation}
where $\omega _{p}=(4\pi ne^{2}/m)^{1/2}$ is the plasma frequency, and 
$\omega _{B}=eB/mc$ is
the electron cyclotron frequency. The positive magnitudes $\Omega _{\perp }$
and $\Omega _{\Vert }$ take into account the Coulomb correlations between
the particles and are expressible via the second frequency moment of the
magnetized plasma conductivity tensor Hermitian part\cite{Adamyan}, so that 
\begin{equation}
\Omega _{\perp }^{2}=\frac{\omega _{p}^{2}}{2}\sum_{\vec{q}\neq 0}S_{ei}(%
\vec{q})\frac{q_{\perp }^{2}}{q^{2}},\quad \Omega _{\Vert }^{2}={\omega
_{p}^{2}}\sum_{\vec{q}\neq 0}S_{ei}(\vec{q})\frac{q_{\Vert }^{2}}{q^{2}},
\label{2.0}
\end{equation}
$S_{ei}(\vec{q})$ being the {partial electron-ion static structure factor},
and $q_{\perp }$ (and $q_{\Vert }$) is the projection of the vector $\vec{q}$
on the direction perpendicular (parallel) to the external magnetic field. An
analysis of these magnitudes is given in the third section, we mention here
only that in the ideal plasma limit both $\Omega _{\perp }$ and $\Omega
_{\Vert }\to 0$. We also wish to emphasize that the electron-ion
correlations is the factor which guarantees the existence of nonvanishing
parameters $\Omega _{\perp }$ and $\Omega _{\Vert }$. Notice that the above
expressions, Eqs. (\ref{ecomp}) for the dielectric tensor components
coincides (within the first order in the ratio $\sqrt{m/M}$) with that
of the quasilocalized charges model developed by Kalman and Golden {\cite
{Kalman}}.

\section{Waves in strongly coupled magnetized plasmas}

If we choose the Cartesian system of coordinates with the z-axis parallel to
the external magnetic field $\vec{B}$, then the {dispersion equation} of
electromagnetic waves propagating in a magnetized plasma takes the form: 
\begin{equation} \label{3.2}
A\,N^{4}\,\,+\,\,B\,N^{2}\,\,+\,\,C\,\,=\,\,0,
\end{equation}
where $N=\omega _{0}/\omega $
is the scalar refraction index, $\omega _{0}=|\vec{k}|c$, and 
\begin{equation}
A=\varepsilon _{\perp }\sin ^{2}\theta \,\,+\,\,\varepsilon _{\Vert }\cos
^{2}\theta ,\quad B=-\,\varepsilon _{\perp }\varepsilon _{\Vert }(1+\cos
^{2}\theta )\,\,-\,\,(\varepsilon _{\perp }^{2}-g^{2})\sin ^{2}\theta ,\quad
C=\varepsilon _{\Vert }(\varepsilon _{\perp }^{2}-g^{2})\qquad ,  \label{3.3}
\end{equation}
and $\theta $ is the angle between the wavevector $\vec{k}$ and the magnetic
field $\vec{B}$.

Eq.(\ref{3.2}) has two different solutions: 
\begin{equation} \label{3.4}
N_{\pm }^{2}=\left[ -B\pm
(B^{2}-4AC)^{1/2}\right] /2A,
\end{equation}
which are usually associated with {ordinary}
and {extraordinary} waves: two different kinds of waves of a given frequency
and with different refraction indices, which can propagate in magnetized
plasmas. These waves are generally elliptically polarized, a wave which
propagates along the external magnetic field is transverse polarized; the
ordinary wave is characterized by the right-handed circular polarization,
the extraordinary wave is left-handed polarized.

The frequencies that satisfy the relation $A(\omega ,\vec{k})=0$ are
traditionally called the {plasma resonances}. Notice that one of the
refraction indices tends to infinity as the frequency approaches the
resonance value $N_{+}^{2}=-B/A$, while the second one remains finite, $%
N_{-}^{2}=-C/B$. A cubic equation with respect to $\omega ^{2}$ can be 
obtained from Eq. (\ref{3.3}). It determines three resonance frequencies. 
This is in contrast to ideal magnetized plasmas 
($\Omega _{\perp }=\Omega _{\Vert }=0$%
), where only two resonances exist (we neglect the ion motion!). For the
case of near longitudinal propagation $\theta \ll 1$ these resonances are 
\begin{equation}
\omega _{\pm }^{(p)}=\frac{1}{2}\left\{ \mp \omega _{B}+\left[ \omega
_{B}^{2}+4\Omega _{\perp }^{2}\right] ^{1/2}\right\} ,\quad \omega _{3}^{p}=%
\sqrt{\omega _{p}^{2}+\Omega _{\Vert }^{2}}\left( 1+\frac{\theta ^{2}}{2}%
\frac{\omega _{p}^{2}\omega _{B}^{2}}{\omega _{p}^{4}-\omega _{p}^{2}\omega
_{B}^{2}-\Omega _{\Vert }^{2}\omega _{B}^{2}}\right) .  \label{res}
\end{equation}

For the case of transverse propagation we found for the refraction indices
poles: 
\begin{equation}
\left( \omega _{\pm }^{(p)}\right) ^{2}=\Omega _{\perp }^{2}+\frac{1}{2}%
\left\{ \omega _{p}^{2}+\omega _{B}^{2}\mp \left[ \left( \omega
_{p}^{2}+\omega _{B}^{2}\right) ^{2}+4\omega _{B}^{2}\Omega _{\perp
}^{2}\right] ^{1/2}\right\} ,\quad \omega _{3}^{(p)}=\Omega _{\Vert }.
\label{rip}
\end{equation}

The zeros of the $N_{\pm }^{2}$ determine the boundaries between the domains
of propagation for different waves. From Eq. (\ref{3.2}) it follows that 
$ N_{\pm }=0$, if the coefficient $C$ is equal to zero. We found three zeros, 
\begin{equation}
\omega _{\pm }^{(0)}=\frac{1}{2}\left\{ \mp \omega _{B}+\left[ \omega
_{B}^{2}+4\left( \omega _{p}^{2}+\Omega _{\perp }^{2}\right) \right]
^{1/2}\right\} ,\quad \omega _{3}^{(0)}=\left( \omega _{p}^{2}+\Omega
_{\Vert }^{2}\right) ^{1/2}.  \label{zeros}
\end{equation}

With the poles and zeros determined, and taking into account that $N_{\pm
}^{2}(\omega =0)=1+\omega _{p}^{2}/\Omega _{\perp }\equiv N_{0}^{2}$, and
the relation $N_{\pm }^{2}(\omega \to \infty )\,\to \,1$, the refraction
indices can be plotted. In Fig.1 the frequency dependence of the refractive
indices for an angle $0<\theta <\pi $ is shown.




The branches of propagation ($N^{2}(\omega )>0$) are associated with the
eigenfrequencies $\omega _{k}$. The latter are given in Fig.2 vs.
wavevector. The modes $\omega _{k}$ are determined by Eq. (\ref{3.2}). Since
in the present approximation Eq. (\ref{3.2}) is the fifth order equation
(with respect to $\omega ^{2}$), we find five eigenmodes. In ideal plasmas
in neglect of the Alfv{\'{e}}n wave only four eigenfrequencies can be found.
From Fig. 2 we observe that in the case of a strongly coupled plasma the
ideal plasma helicon wave splits into two branches, which we call the
strongly coupled plasma whistling sound waves.

Thus in strongly coupled magnetized plasma there can exist five eigenmodes:
ordinary and extraordinary whistling sound waves, the slow extraordinary,
the ordinary and the fast extraordinary waves.

Consider now in more details the dispersion of the whistling sound waves at
small wavenumbers, i.e., when $\omega \ll \omega _{B}$. In this spectral
region the dispersion equation reduces to a quadratic equation with respect 
to $%
\omega ^{2}$. For the case of parallel with respect to the external magnetic
field propagation the corresponding solution reads:

\begin{equation}  \label{soundwave}
\omega_k^{(4,5)}=\frac{1}{2} \left\{ \pm \, \frac{\omega_B \omega_0^2} {%
\omega_p^2 + \Omega_{\perp}^2 + \omega_0^2} + \left[ \frac{\omega_B^2
\omega_0^4}{(\omega_p^2 + \Omega_{\perp}^2 + \omega_0^2)^2} + 4\, \frac{
\omega_0^2 \Omega_{\perp}^2}{\omega_p^2 + \Omega_{\perp}^2 + \omega_0^2}
\right]^{1/2} \right\} \quad.
\end{equation}

In ideal magnetized plasmas (i.e., when $\Omega _{\perp }=0$) the solution
of Eq.(\ref{soundwave}) represents then the spiral wave - the {helicon},
or the whistler, the frequency of which equals {\cite{A}} 
\begin{equation}
\omega _{k}^{(h)}=\frac{\omega _{0}^{2}\omega _{B}}{(\omega _{p}^{2}+\omega
_{0}^{2})},  \label{3.15}
\end{equation}
and tends to zero as $|\vec{B}|\to 0$.

For the case of strong interaction between the particles and at small
wavenumbers, i.e. when $\Omega _{\perp }>(\omega _{B}\omega _{0})/\omega
_{p} $, the solutions of Eq.(\ref{soundwave}) describe the ordinary and
extraordinary whistling sound waves propagating in strongly coupled plasmas,

\begin{equation}
\omega _{k}^{(4,5)}=v_{s}k\pm \frac{1}{2}\,\frac{\omega _{B}\omega _{0}^{2}\
\omega _{p}^{2}}{(\omega _{p}^{2}+\Omega _{\perp }^{2})^{2}}\quad ,
\label{twosound}
\end{equation}
with the whistling sound velocity $v_{s}=c\Omega _{\perp }/\sqrt{\omega
_{p}^{2}+\Omega _{\perp }^{2}}.$

The parameter $\Omega _{\perp }$ will be estimated in the next section.
\section{Estimate of the frequencies $\Omega _{\Vert }$ and $\Omega _{\bot }$%
.}

For our purposes it is sufficient to make a simple estimate of the
frequencies $\Omega _{\Vert }$ and $\Omega _{\bot }$ without including their
magnetic field dependence, within the random-phase approximation (RPA), and
in the hydrogen plasma model. In this approximation they coincide 
\begin{equation} \label{Omegaparam}
\Omega _{\Vert }^{2}=\Omega _{\bot }^{2}=h_{ei}\left( 0\right) \, \omega_p^2 
/3= \omega_p^2 \, \sum\nolimits_{\vec{q}\neq 0}S_{ei}(q)/3 \, ,
\end{equation}
and are directly related to \cite
{Adamyan} the zero separation value of the electron-ion correlation
function, $h_{ei}\left( 0\right) .$ The electon-ion structure factor can be
estimated in a Coulomb system as \cite{agd}:

\begin{equation}
S_{ei}(q)=\frac{4\pi e^{2}}{n\beta q^{2}}\frac{\Pi _{e}(q)\Pi _{i}(q)}{%
\varepsilon (q)}.  \label{sei}
\end{equation}

Here $\Pi _{e}(q),\Pi _{i}(q)$, and $\varepsilon (q)$ are the static
electronic and ionic polarization operators (real parts), and the dielectric 
function,
respectively,
$\beta ^{-1}$ is the system temperature in energy units.
The ions can be considered as classical particles, and we put $\Pi
_{i}(q)=n\beta .$ For $\Pi _{e}(q)$ we employ a rational interpolation \cite
{tvt} 
\begin{equation}
\Pi _{e}(q)=\frac{\gamma ^{4}/4\pi e^{2}}{q^{2}+\gamma ^{4}\lambda _{D}^{2}},
\label{pie}
\end{equation}
constructed to satisfy both long- and short-wavelength limiting conditions
of the RPA: $\lambda _{D}^{2}=4\pi e^{2}n\beta $ and $\gamma ^{4}=16\pi
ne^{2}m/\hbar ^{2}.$

After a straightforward calculation we obtained: 

\begin{equation}
h_{ei}\left( 0\right) =2\alpha r_{s} 
{\frac{\sqrt{2}}{\sqrt{B+\sqrt{QS}}+\sqrt{B-\sqrt{QS}}},\text{ if }Q>0 
\atopwithdelims\{. \frac{1}{\sqrt{S}},\text{\qquad \qquad \qquad \thinspace 
\quad \quad if }Q\leq 0}
,  \label{Q}
\end{equation}

where $S=B+\sqrt{2A/3}$, and $Q=B-\sqrt{2A/3}$, $A=4\alpha r_{s}/\pi $, $%
\alpha =\left( 4/9\pi \right) ^{1/3}=\allowbreak 0.\,521$, and $B=\left( \pi
/3\right) ^{1/3}\left( A/4\Gamma \right) +A/6\Theta .$ Usual notations are
introduced here: $k_{F}$ is the Fermi wavenumber, $r_{s}=a/a_{B}$ is the
Brueckner parameter, i.e., the Wigner-Seitz distance $a$ in the units of the
Bohr radius, $\Gamma =\beta e^{2}/a$, and $\Theta =\left( \beta E_{F}\right)
^{-1}$, $E_{F}$ being the Fermi energy. Notice that $r_{s}=\Gamma \Theta
/0.543.$

In the case of weakly coupled plasmas with $\Gamma \longrightarrow 0$,

\begin{equation}
h_{ei}\left( 0\right) \simeq \allowbreak 3.\,4\Gamma \sqrt{\Theta } \simeq 
\allowbreak 13. \,1 \sqrt{\frac{eV}{T}}\,,
\label{hei0g0}
\end{equation}
with the temperature T measured in units of eV.

Eq.(\ref{Q}) (or Eq.(\ref{hei0g0}) in the limit of weak coupling) together 
with Eq.(\ref{Omegaparam}) determine the magnitudes $\Omega _{\perp }$ and 
$\Omega _{\Vert }$.
\section{Conclusions}

In this note the dispersion laws for electromagnetic waves in cold
magnetized plasmas are analyzed. Our analysis is based on the expression for
the plasma dielectric tensor obtained from the classical theory of moments
without using perturbation parameters. Thus both the cases of weak and
strong Coulomb coupling are regarded. A qualitative distinction between
systems with weak and strong Coulomb coupling is established in their
low-frequency electromagnetic wave propagation spectra. It is shown that the
weakly coupled plasma helicon branch splits in strongly coupled plasmas into
two whistling sound branches. The coupling parameters thermodynamic
dependence is estimated.

{\bf Acknowledgments. }This work was partly financed by the Deutsche
Forschungsgemeinschaft and the Polytechnic University of Valencia, Spain.

\newpage

\begin{center}
{\bf FIGURE CAPTIONS}
\end{center}

\begin{description}

\item[(Figure 1)] Squares of  refraction indices of strongly 
coupled magnetized plasma vs. frequency (in arbitrary units) 
1-fast extraordinary wave; 2- ordinary wave; 3 - slow extraordinary wave; 
4,5 - strongly coupled plasma whistling sound waves; ($0 < \theta < \pi$).

\item[(Figure 2)] Frequencies of various eigenmodes of strongly coupled and 
ideal magnetized plasma vs. wavevector (in arbitrary units) 1-5 see Fig.1; 
3' - slow extraordinary wave of ideal plasma;  4' - helicon wave of ideal 
plasmas  ($0 < \theta < \pi$).

\end{description}

\newpage

Figure 1. (Electromagnetic modes in cold magnetized strongly coupled plasmas
authors: Tkachenko, Ortner, Rylyuk)

\begin {figure} [h] 
\unitlength1mm
  \begin{picture}(170,160)
\put (0,-10){\epsfig{figure=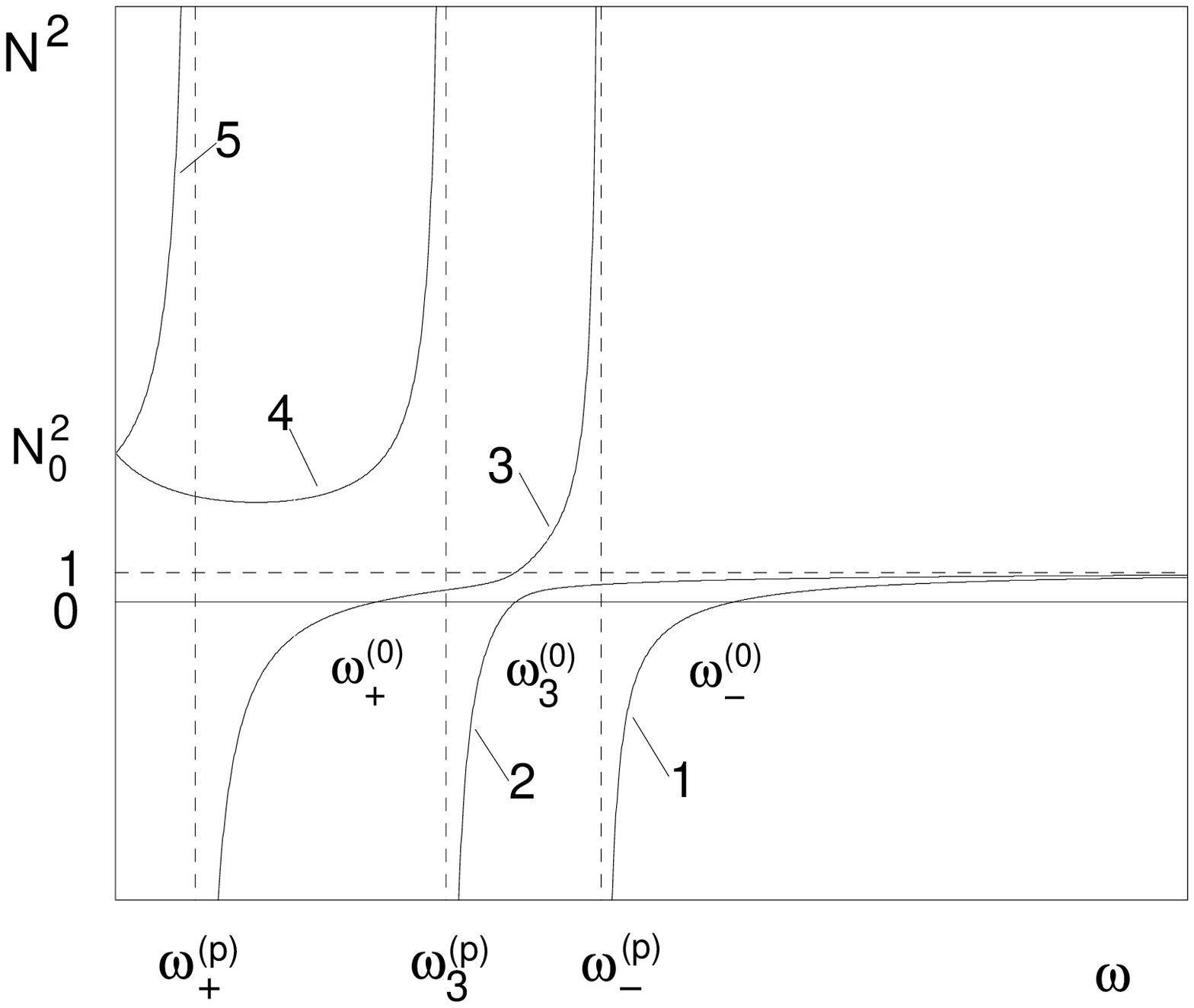,width=15.0cm,height=14.0cm,angle=0}}
 \end{picture}\par
\end{figure}

\newpage

Figure 2.  (Electromagnetic modes in cold magnetized strongly coupled plasmas
authors: Tkachenko, Ortner, Rylyuk)

\begin {figure} [h] 
\unitlength1mm
  \begin{picture}(170,160)
\put (0,-10){\epsfig{figure=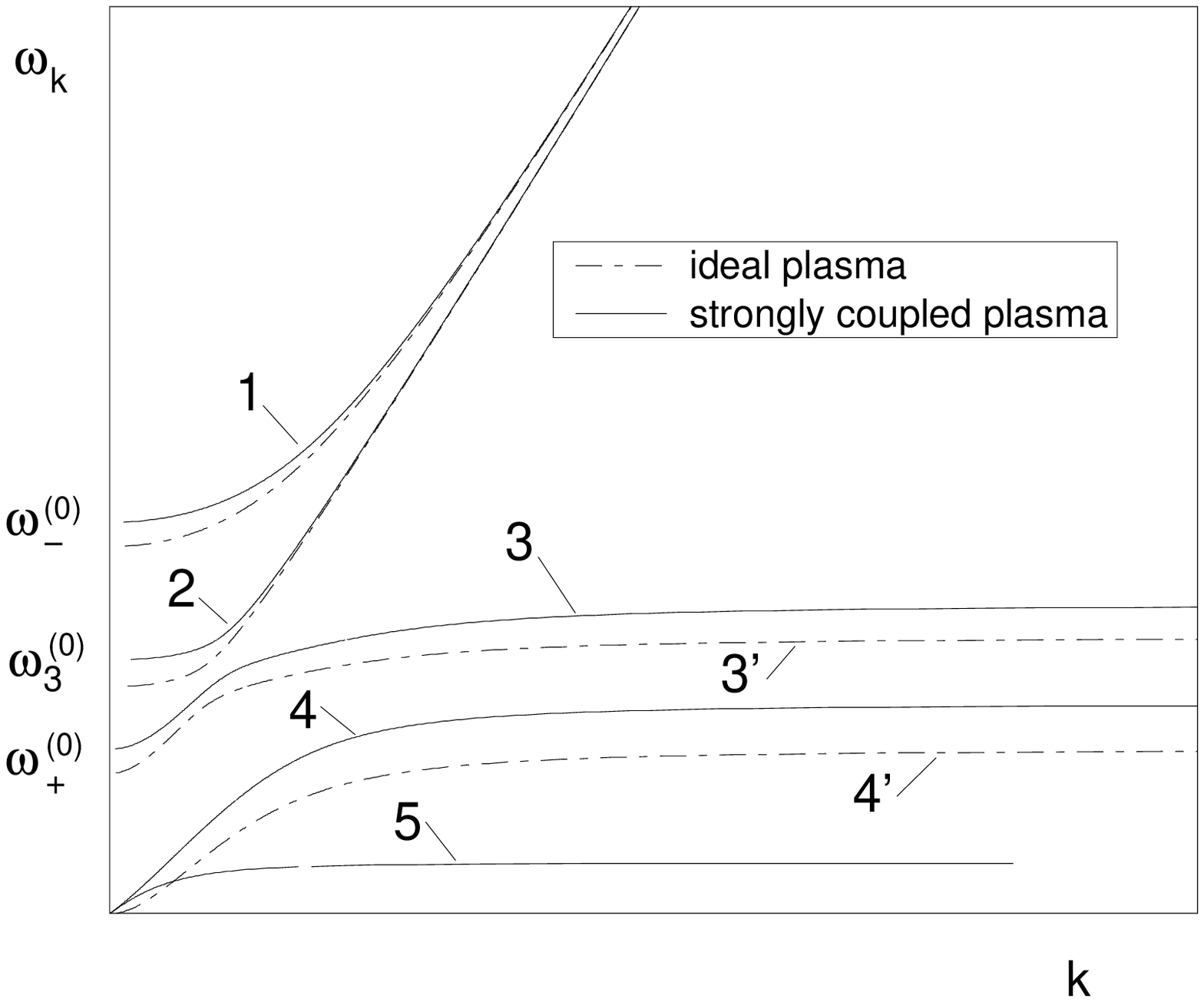,width=15.0cm,height=14.0cm,angle=0}}
 \end{picture}\par 
\end{figure}

\end{document}